\documentclass[aps,prb,preprint,onecolumn,showpacs,unsortedaddress,showkeys,preprintnumbers,amsmath,amssymb,nofootinbib]{revtex4}
\usepackage{graphicx}
\usepackage{dcolumn}
\usepackage{bm}

\begin{document}

\title{Numerical simulations versus theoretical predictions for a non-Gaussian noise induced escape problem in application to full counting statistics}

\author{I. A. Khovanov$^{1,2}$}
\email{i.khovanov@warwick.ac.uk}
\author{N. A. Khovanova$^1$}
\affiliation{$^1$ School of Engineering, University of Warwick, Coventry CV4 7AL, United Kingdom}
\affiliation{$^2$ Centre for Scientific Computing, University of Warwick, Coventry, CV4 7AL, United Kingdom}

\begin{abstract} 
A theoretical approach for characterising the influence of asymmetry of noise distribution on the  escape rate of a multi-stable system is presented. This was carried out via the estimation of an action, which is defined as an exponential factor in the escape rate, and discussed in the context of full counting statistics paradigm. The  approach takes into account all cumulants of the noise distribution and demonstrates an excellent agreement with the results of numerical simulations. An approximation of the third order cumulant was shown to have limitations on the range of dynamic stochastic system parameters. The applicability of the theoretical approaches developed so far is discussed for an adequate characterisation of the escape rate measured in experiments.   
\end{abstract}


\maketitle

\section{Introduction}

{\it Shot noise}~\cite{Levitov:03} 
characterises transport properties of mesoscopic conductors. Therefore, studying properties of the shot noise is essential for understanding the behaviour of the mesoscopic carries. The properties can be described by the full counting statistics approach \cite{Levitov:96} which considers {\it the third and higher order} cumulants, also known as irreducible correlators. At the same time first and second cumulants specify  properties of equilibrium {\it symmetrical} Johnson-Nyquist noise which is different from the shot noise. Typically~\cite{Levitov:03}, the statistics of the shot noise are characterised by a non-symmetrical distribution, e.g. binomial or Poisson. A scheme for a qualitative characterisation of the distribution asymmetry {\it via} measurements of the  escape rate of an auxiliary multistable system driven by fluctuations has  been  recently suggested~\cite{Tobiska:04,Pekola:04}. The main idea was to study the escape rate of a {\it noise detector}  for characterisation of acting fluctuations which are the output of a mesoscopic system. This scheme was extensively discussed in a number of theoretical papers~\cite{Ankerhold:07,Grabert:08,Novotny:09} and implemented experimentally with the Josephson junction as a noise detector  driven by Poisson noise~\cite{Timofeev:07,Masne:09}. A combination of theoretical, numerical and experimental investigations was presented~\cite{Huard:07} showing some correspondences and disagreements between the theory, numerics and experiment. However, despite the progress made in the theoretical description of the shot-noise and, in turn, properties of the noise-detector scheme, certain questions still remain. For example, accuracy of the suggested theoretical approaches were not validated by numerical simulations. There are also certain disagreements between several published theoretical approaches~\cite{Novotny:09} thus resulting in controversy~\cite{Sukhorukov:10}. Notably, the reports indicate poor correspondence between the theoretical and experimental results~\cite{Huard:07,Novotny:09} and several outstanding issues have been identified~\cite{Novotny:09}. In particular, the literature cited two important aspects: (i) validity of the use of a third order cumulant approximation in theoretical approaches and (ii) omission of the prefactor in the expression of the escape rate~\cite{Novotny:09}. There is, however,  reported research~\cite{Novotny:09} which places the scheme into a strong {\it nonlinear} regime for maximising distribution asymmetry. 

In the majority of published articles a similar theoretical model of the noise detector scheme has been used that corresponds to a non-linear oscillator driven by a mixture of white Gaussian and Poisson noises. The presence of the Poissonian process leads to asymmetry of the noise distribution.  The characterisation of the degree of this asymmetry has become the focus of such theoretical considerations. To make the model analytically tractable, an approximation of {\it the third order cumulant} is often applied together with a high barrier assumption~\cite{Ankerhold:07,Grabert:08,Novotny:09}. If the latter corresponds to a typical experimental set~\cite{Timofeev:07,Masne:09} the former places limitations on the theoretical predictions, and the degree of the limitations is still not well understood. Several theoretical approaches~\cite{Pekola:04,Sukhorukov:07,Sukhorukov:10} assume a high friction limit resulting in over-damped model dynamics, whereas weak damping (under-damped)  dynamics is experimentally observed~\cite{Timofeev:07,Masne:09}. 

A revision of the existing approaches explained earlier is necessary in order to shed light on the
non-Gaussian noise induced escape problem. The aim of this research is therefore to present a generic approach beyond the third cumulant limit, and to compare the theoretical predictions with the results of numerical simulations. The revision in the present work follows the approaches adopted in previous studies~\cite{Grabert:08,Novotny:09} demonstrating the presence of constraints on the noise characteristics when theoretical predictions are compared with numerical and/or experimental results. We also show that the absence of a prefactor in the estimation of the escape rate may result in a large error in theoretical predictions.

Theoretical considerations~\cite{Ankerhold:07,Huard:07,Sukhorukov:07,Grabert:08,Novotny:09}, as well as the present work, are motivated by the use of the escape rate for characterisation of shot noise in mesoscopic conductors. However, the general theoretical setting is applicable for a wider range of problems including vibrations in civil structures \cite{Casciati:90}, switching in MEMS and NEMS \cite{Khovanova:11,Dykman:12} devices,  neuronal dynamics \cite{Richardson:05} and ion channel permittivity \cite{Schroeder:09,Kaufman:13}.  It is important to mention that noise-induced escape and the corresponding mean first-passage time problem in the presence of non-Gaussian noise (also known as white shot noise) have not been discussed as comprehensively as white or coloured Gaussian noise \cite{Hanggi:05}. Nonetheless, the literature is extensive and nearly all publications have dealt with one-dimensional potential systems~\cite{Masoliver:87,Masoliver:93,Kim:07,Grigoriu:04,Dykman:10}    for the overdamped case. In some cases, additional limitations on noise characteristics  \cite{Masoliver:87,Masoliver:93,Grigoriu:04} allowed  the problem to be solved analytically and made the task analytically intractable for other cases \cite{Masoliver:93}.

The approach adapted in the present work is based on the practical realisations of a noise detector used for full counting statistics~\cite{Timofeev:07,Masne:09}. This approach provides an experimental basis for studying the escape problem in the presence of non-Gaussian noise. The dynamic behaviour of the detector is under-damped and the acting noise is white (uncorrelated) with a finite second cumulant; there are no explicit limitations on the shape of the noise distribution. This experimental setting is quite broad involving wide ranging research applications as mentioned above, and thus extends beyond the detection of noise statistics.

Dynamic system, noise properties as well as the theoretical approach for the estimation of the escape rate for non-Gaussian noise are presented in Sec.\ \ref{sec:theory}. We compare theoretical and numerical results in Sec.\ \ref{sec:numerics} and discuss applicability of the third cumulant limit in Sec.~\ref{sec:3d}. The main conclusions are summarised in Sec.\ \ref{sec:concs}.

\begin{figure}[t!]
\includegraphics[width=8cm]{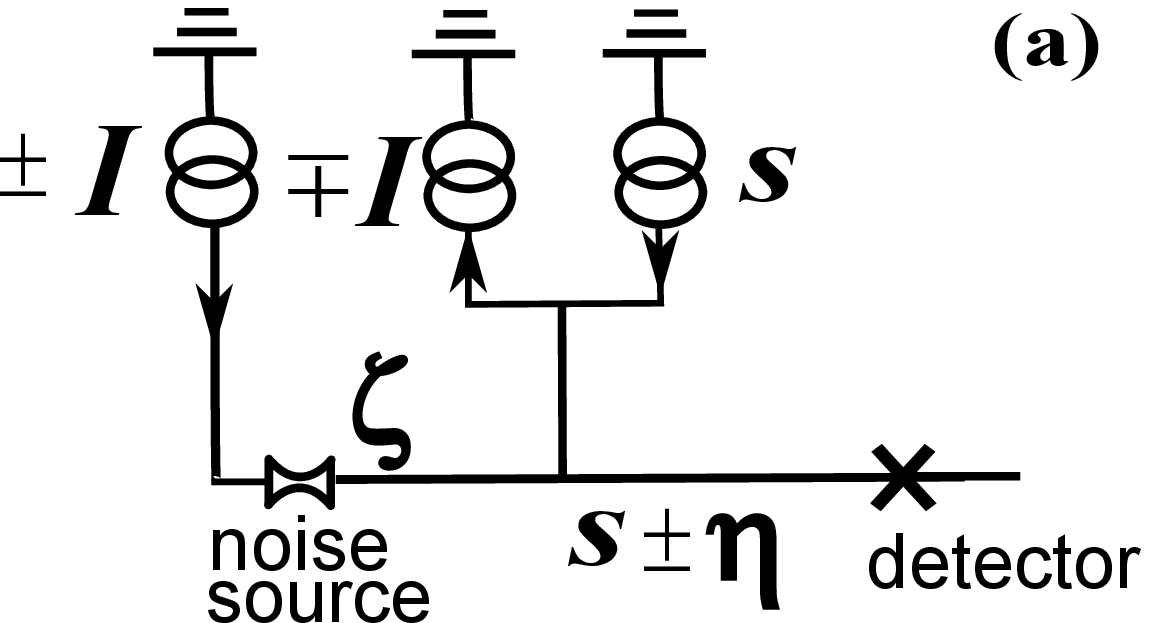}\\
\includegraphics[width=8cm]{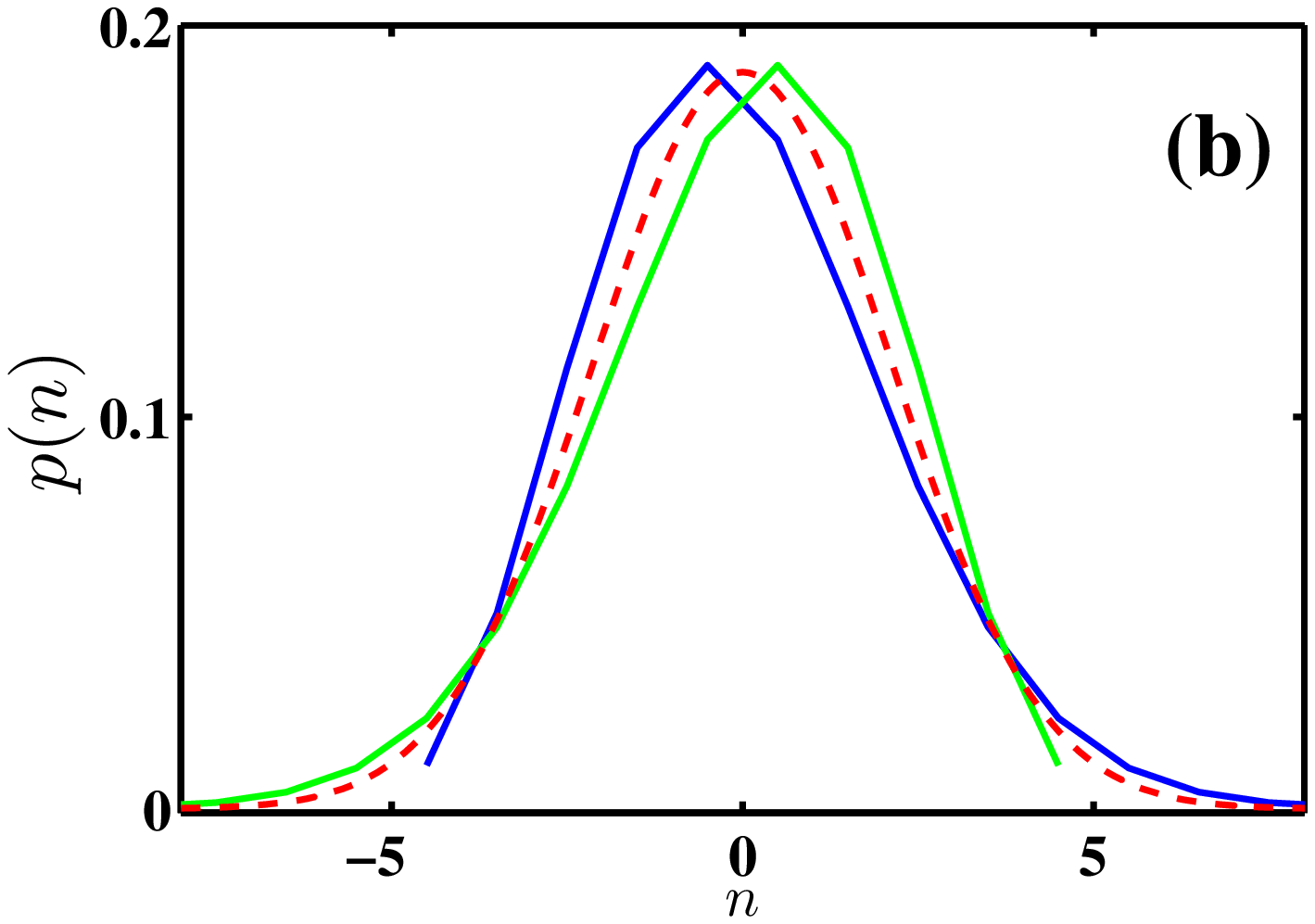}\\
\includegraphics[width=8cm]{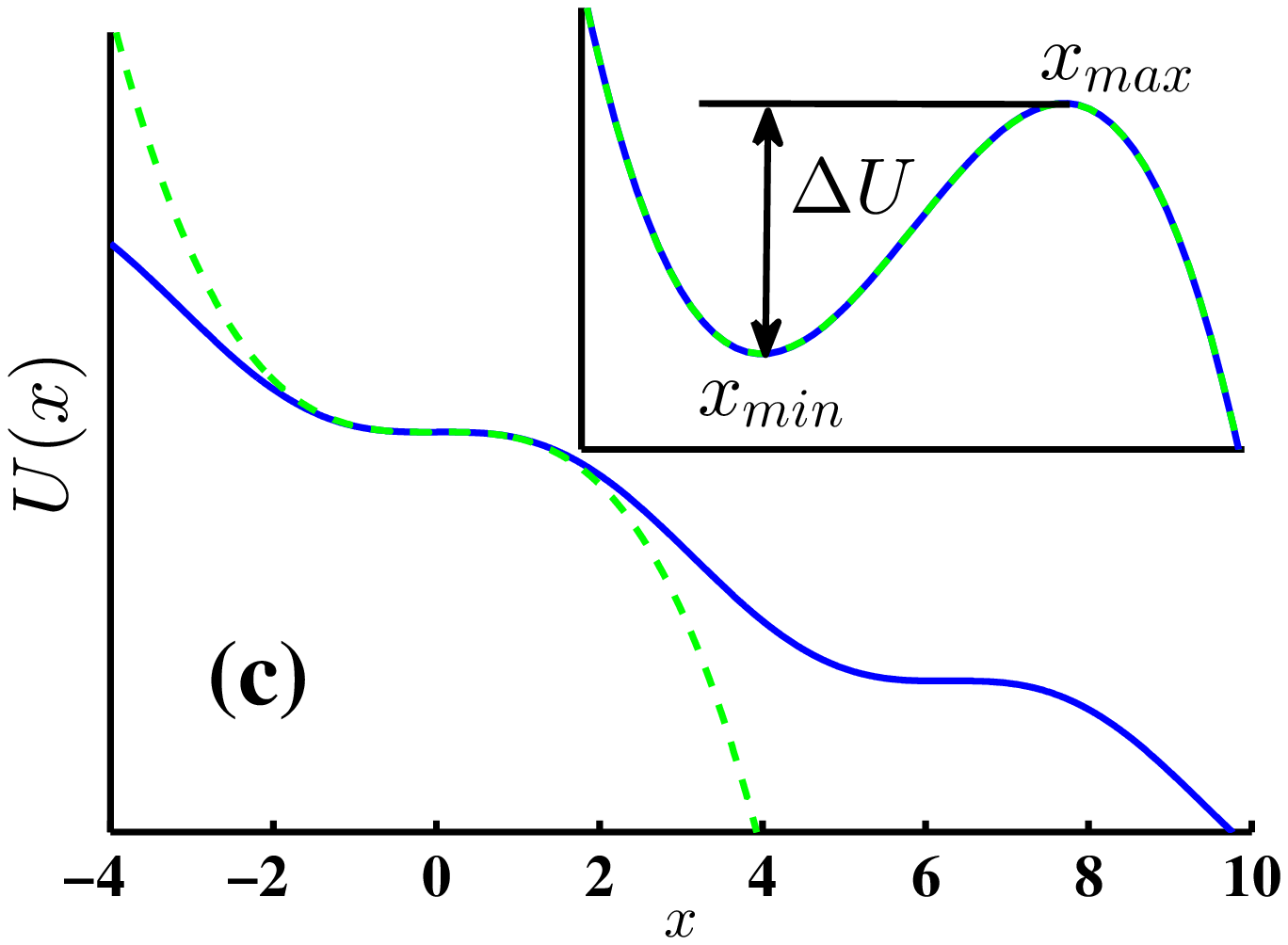}
\caption{\label{fig1} (color online) (a) Sketch of experimental setup. Symbols $I$ and $s$ denote biased currents. Direction of the currents are shown by arrows. (b) Schematic of the asymmetry of noise distribution. Poisson distributions (\ref{p_dist}), initial and flipped, with $\Lambda=4.5$ and $\Delta t=1$ are shown by solid (blue and green) curves, and  Gaussian distribution with a zero mean value and a standard deviation equal to $\Lambda$ was used to draw dashed (red) curve. (c) The tilted $U(x)=\sin(x)-sx$ and third order $U(x)=ax-bx^3/3$ potentials are shown by solid (blue) and dashed (green) curves respectively with the following parameters: $s=0.99065$, $a=0.0093101$, $b=0.4965$. The inset shows the part of the potentials in the vicinities of their minimal and maximal values (first extrema for the tilted potential) which are undistinguished.
}
\end{figure}

\section{Dynamic system and theoretical approach}
\label{sec:theory}

A simplified experimental scheme, presented in Fig.~\ref{fig1}~(a), shows a noise source and a threshold detector, both biased by current and based on the Josephson junction~\cite{Timofeev:07,Huard:07,Masne:09}. For this system, the voltage drop in the noise source is larger than thermal fluctuations.  Tunnelling events are a dominant source of carriers implying that the noise source is in a pure {\it shot noise} regime. An additional current bias (middle of Fig.~\ref{fig1}~(a)) is used to remove a constant component $I$ of the shot noise $\zeta(t)$ indicating that {\it zero-mean} shot noise $\eta(t)$ is acting on the detector. In contrast, the detector is in the thermal regime as the voltage drop of the detector is smaller than thermal fluctuations. It is noted that the experiments~\cite{Timofeev:07,Huard:07,Masne:09} are carried out in a low temperature environment with a minimal temperature of around 20 mK. For such low temperature values the quantum effects in the detector should be strong as predicted by the theory (for example reference~\cite{Eckern:84}). However, previous experimental results~\cite{Pekola:05,Timofeev:07,Huard:07,Masne:09,Kivioja:05} state that the dynamics of the detector can also be described within the classical limit. Another potentially important factor refers to the presence of the feedback effect of the detector to the noise source~\cite{Kindermann:04,Urban:09}. This factor can be neglected in our consideration due to the  feedback being considered small in the reported experiments~\cite{Timofeev:07,Huard:07,Masne:09,Novotny:09}.

The Josephson junction as a noise source produces shot noise with Poissonian statistics~\cite{Khlus:87,Lesovik:89,Blatner:00,Levitov:03}. Poisson noise is known as a rare event process with an asymmetric probability distribution~\cite{Feller_Book}. The presence of the asymmetry  means that depending on the sign of the currents $I$ (left and middle current sources in Fig.~\ref{fig1}~(a)) the right and left tails of the noise probability distribution are of different widths (Fig.~\ref{fig1}~(b)). This asymmetry is reflected in the difference (asymmetry)  of escape rates in the detector that has a multistable potential (Fig.~\ref{fig1}~(c)). Thus, the difference between the escape rates characterises the asymmetry of the noise distribution and, consequently, the degree of non-Gaussianity. In order to extract quantitative information on non-Gaussianity from the experimental measurements, a theoretical description (in the form of  a mathematical  expression in the simplest case) is required that conjoins the escape rate with the asymmetry parameter(s). So far the third cumulant, as mentioned above, has been considered as the main asymmetry parameter, although all cumulants of higher than the second order contribute~\cite{Cramer_Book,Malakhov_Book}. Note that the values of the first and second cumulants as well as detector parameters can be defined by using well established measurement techniques~\cite{Huard:07,Kivioja:05}, and therefore these values are considered as 'known'. The validity of the theoretical prediction is crucial for the entire experimental approach.

The dynamic system under consideration, which describes the experimental setup \cite{Huard:07,Masne:09,Novotny:09}, is as follows
\begin{eqnarray}
\label{sin_sys}
\ddot{x}+\alpha \dot{x}+\frac{dU}{dx} =\sqrt{\alpha D} \xi(t) \pm \eta (t) \  \\
\label{cnoise}
\eta(t)=\zeta(t)-I \ .
\end{eqnarray}
Equation (\ref{sin_sys}) is in a dimensionless form and all the parameters are normalised. An equivalent non-normalised equation with corresponding values of parameters can be found in published papers~\cite{Huard:07,Novotny:09}. Normalised coordinate $x$ corresponds to the current of the Josephson junction with biased potential $U(x)=\sin(x)-sx$ and damping coefficient $\alpha$. The bias $s$ is selected in such a way that a tilted multistable (washboard) potential is formed in the system,  and  a noise-induced escape from one of the stable states via the lowest potential barrier (Fig.~\ref{fig1}~(c)) is analysed. In case of a tilted form, the system (\ref{sin_sys}), after escape, evolves along the potential (so-called running mode) and this motion can be easily detected in experiments~\cite{Huard:07,Masne:09}.
Noise $\xi(t)$ corresponds to the thermal noise in the detector. It is Gaussian white noise with unit variance and zero mean value: $\langle \xi(t) \rangle =0$, $\langle \xi(t)\xi(0) \rangle =\delta(t)$; $D$ defines noise intensity. 
Term $\eta(t)$ in (\ref{sin_sys}) describes a shot noise of the source and consists of two components (\ref{cnoise}):  $I$ is bias current  applied to the source and $\zeta(t)$ corresponds to the Poisson white noise which can be represented as a sum of independent pulses \cite{Klyackin_Book}
\begin{equation}
\label{shot_noise}
\zeta(t) = \sum_{i=1}^{N} z_i g(t-t_i).
\end{equation}
In expression~(\ref{shot_noise}), $z_i$ are independent random amplitudes of pulses, function $g(t-t_i)$ describes the pulse shape, $t_i$ are independent random times of pulse appearance. Time intervals $\tau_i=t_{i+1}-t_i$ between two subsequent pulses has exponential distribution 
\begin{equation}
\label{exp_dist}
p(\tau_i)=\Lambda \exp(-\Lambda \tau_i)
\end{equation}
where $\Lambda$ is a parameter of the Poisson noise and defines the frequency of the events.
The number of pulses $n$ within the time interval $\Delta t$ follows a Poisson distribution
\begin{equation}
\label{p_dist}
p(n)=\frac{(\Delta t \Lambda)^n \exp(-\Delta t \Lambda )}{n!}.
\end{equation}
Following previous approaches \cite{Ankerhold:07,Huard:07,Grabert:08,Novotny:09} we consider $\delta$-impulses of the same amplitude $\lambda$:
\begin{equation}
\label{poi_noise}
\zeta(t) = \sum_{i=1}^{N} \lambda \delta(t-t_i).
\end{equation}
The Poisson noise (\ref{poi_noise}) is characterised by an infinite number of non-zero cumulants defined as
\begin{equation}
\label{cumulants}
\chi_s (0, t_1,\ldots t_s) = \Lambda \lambda^s \delta(t_1)\ldots \delta(t_s),
\end{equation}
where $\chi_s$ represents an $s$-order cumulant. Since the first cumulant is non-zero, the noise produces a bias
$\lambda \Lambda$ which is removed by term $I$ in (\ref{cnoise}) (the middle current source in Fig.~\ref{fig1}~(a)), therefore  $I=\lambda \Lambda$. Thus, we consider a zero-mean non-Gaussian noise  $\eta(t)$ acting together with a white noise $\xi(t)$ on the system (\ref{sin_sys}).
All cumulants of $\eta(t)$ are equal to cumulants $\chi_s$ except the first cumulant, which is equal to zero.

The task consists of the estimation of the difference between two mean first-passage times ($T_+$ and $T_-$) corresponding to the opposite signs in front of $\eta(t)$ in (\ref{sin_sys}). Both $T_+$ and $T_-$ are  experimentally measured quantities which lead to an  asymmetry factor \cite{Huard:07,Novotny:09} 
\begin{equation}
\label{Assymt}
\Gamma_{T} = \frac{T_+}{T_-}-1 .
\end{equation}
The value of $\Gamma_{T}$ characterizes the asymmetry of the noise distribution and the lower index $T$ is used to stress that the factor is derived from measurements of times $T_{\pm}$. 
 Due to escape having {\it an activation character}, the times $T_{\pm}$ can be expressed in the following form~\cite{Hanggi:05}:
\begin{equation}
\label{Kramers1}
T_{\pm} = Z_{\pm} \exp \left( \frac{S_{\pm}}{\theta} \right) ,
\end{equation}
where $Z_{\pm}$ and $S_{\pm}$ are the prefactor and action respectively, $\theta$ is an effective intensity of both Gaussian and non-Gaussian noise in (\ref{sin_sys}).  
 Further, it is implicitly assumed \cite{Ankerhold:07,Huard:07,Sukhorukov:07,Grabert:08,Novotny:09} that prefactor $Z_{\pm}$ can be omitted leading to the asymmetry factor in the form
\begin{equation}
\label{Assym1}
\Gamma_{S}  =\exp \left( \frac{S_+ - S_-}{\theta} \right)-1 \ .
\end{equation}
The subscript $S$ in~(\ref{Assym1}) indicates that we need to measure actions $S_{\pm}$ rather than times $T_{\pm}$. Omission of the prefactors $Z_{\pm}$ is equivalent to the assumption that $Z_+=Z_-$. The validity of this assumption was not verified and we discuss this matter below.
Thus, the actions $S_{\pm}$ are the subject of theoretical approaches, whereas the times $T_{\pm}$ are measured experimentally. This indicates that the actions rather than $T_{\pm}$ must be extracted from experiments or numerical simulations  for the proper use of the theory. Note, that this aspect has not been addressed in published papers.

Although all previously suggested theoretical approaches \cite{Ankerhold:07,Huard:07,Sukhorukov:07,Grabert:08,Novotny:09} are very similar, there is no unified framework to follow. Therefore, we suggest our version which is based on the approaches described in publications~\cite{Novotny:09,Grabert:08}. This will enable us to compare two approaches: the first approach with only third cumulant taken into account and the second approach with all cumulants (\ref{cumulants})  considered. In this way the importance of higher order cumulants will be investigated. 

The starting point of the theoretical development is the Fokker-Plank equation (FPE) corresponding to the Langevin equation (\ref{sin_sys})~\cite{Novotny:09}  (see articles\cite{Klyackin_Book,Denisov:09} for details of the derivation of the term describing non-Gaussian noise)
\begin{eqnarray}
\label{FPE}
&&\frac{\partial P}{\partial t} = - \frac{\partial}{\partial x}\left(y P \right) -
\frac{\partial}{\partial y}\left[ \left( -\alpha y -\frac{d U(x)}{d x} \right) P \right]+ \nonumber \\
&&\frac{1}{2}\alpha D \frac{\partial^2}{\partial y^2}P + \Lambda \left[
\exp\left(\mp \lambda \frac{\partial}{\partial y} \right) -1 \right]P \ .
\end{eqnarray}
In equation (\ref{FPE}) a new variable $y=\dot{x}$ is introduced and $P\equiv P(x,y,t)$ is probability density. The exponent in the last term in (\ref{FPE}) describes Poisson noise and has the following Maclaurin series representation \cite{AbramowitzStegun}:
\begin{eqnarray}
\label{expP}
\exp\left(\mp \lambda \frac{\partial}{\partial y} \right)= \sum_{j=0}^\infty \frac{(\mp \lambda)^j \frac{\partial^j}{\partial y^j} }{j!} .
\end{eqnarray}
Let us consider the solution of equation (\ref{FPE}) in exponential form~\cite{Novotny:09}, that is 
\begin{eqnarray}
\label{expsc}
P \propto \exp\left( \frac{S}{\theta} \right) \ ,
\end{eqnarray}
with action $S$ and the effective noise intensity $\theta$. Note that an exponential form has been used in all recent theoretical approaches \cite{Ankerhold:07,Huard:07,Sukhorukov:07,Grabert:08,Novotny:09}. However, particular forms of $\theta$ are varied from one approach to another. Importantly, the proportionality symbol is used in (\ref{expsc}) because equality  would require an additional prefactor $Z$. 

Effective noise intensity $\theta$ is an asymptotic parameter of the problem. Assuming that higher cumulants of a non-Gaussian noise are smaller than the second cumulant,  $\theta$ can be chosen \cite{Novotny:09} in the form $\theta= \alpha D +\lambda^2 \Lambda$, i.e. $\theta$ is proportional to the second moment of the sum of the Gaussian and non-Gaussian noises.  Note that the second moment is a quantity measured experimentally. The selection of $\theta$ and the form of (\ref{expsc}) require an additional verification, which is performed later.

Following the Wentzel-Kramers-Brillouin (WKB) approximation \cite{Novotny:09,FreidlinWentzel}, we substitute (\ref{expsc}) into the FPE (\ref{FPE}) and keep the leading order ($1/\theta$) terms only. The latter formally means the use of the zero-noise limit as $\theta \rightarrow 0$. The final result can be written as the following Hamiltonian system of equations 
\begin{eqnarray}
\label{Ham_sys}
\begin{aligned}
\dot{x} &=y  \\
\dot{y} &= - \alpha y- \frac{d U}{d x} -\frac{\alpha D}{\theta}p_y \pm \lambda \Lambda \left[ \exp\left( \frac{\mp \lambda p_y}{\theta}\right)-1\right]  \\
\dot{p}_x &= p_x \frac{d^2 U}{d x^2}  \\
\dot{p}_y &= -p_x+\alpha p_y  
\end{aligned} .
\end{eqnarray}
In (\ref{Ham_sys}), $p_x\equiv \frac {\partial S}{\partial x}$ and $p_y\equiv \frac{\partial S}{\partial y}$ are conjugated moments. In contrast to the Gaussian case \cite{FreidlinWentzel}, the asymptotic parameter $\theta$ is not eliminated; all the parameters characterising noise also present in the final system of equations (\ref{Ham_sys}). The approach, however, can be only applied formally for $S \gg \theta$. Below we return to this point.

System (\ref{Ham_sys}) has to be completed by two boundary conditions corresponding to a transition from the minimum of the potential $(x=x_{min},y=0)$ to its maximum $(x=x_{max},y=0)$ (inset in Fig.~\ref{fig1}~(c)). These boundary conditions~\cite{FreidlinWentzel} specify a heteroclinic (connecting two saddle states) trajectory of system (\ref{Ham_sys}) and these are the following 
\begin{eqnarray}
\label{bconds}
\begin{aligned}
\mbox{for } t_i\rightarrow -\infty: x=x_{min},  y=0, p_x=0, p_y=0 \\
\mbox{for } t_f\rightarrow  \infty: x=x_{max},  y=0, p_x=0, p_y=0; \\
\end{aligned} 
\end{eqnarray}
where $t_i$ and $t_f$ are the initial and final time moments.
If the solution of the boundary problem exists \cite{FreidlinWentzel}, it can be used to calculate the action difference $S_{max}-S_{min}$ corresponding to the minimal energy required for the system to migrate from the bottom, $x_{min}$, to the top potential, $x_{max}$.  We assume that $S_{min}=0$ and therefore $S=S_{max}$. Denoting coordinates of a heteroclinic trajectory as $(\tilde{x},\tilde{y},\tilde{p_x},\tilde{p_y})$, the action is determined by the following expression
\begin{eqnarray}
\label{action}
S_{\pm}=&\int_{t_i}^{t_f} dt\left\{ -\frac{\alpha D}{2\theta}\tilde{p}_y^2+\theta \Lambda \left[ \exp\left( \frac{\mp \lambda \tilde{p}_y}{\theta}\right)-1 \pm \frac{\lambda}{\theta}\tilde{p}_y \right] \pm \right. \nonumber \\
 &\left. \tilde{p}_y \lambda \Lambda \left[ \exp\left( \frac{\mp \lambda \tilde{p}_y}{\theta}\right)-1 \right] 
 \right\} \ 
\end{eqnarray}
where $t_i$ and $t_f$ are initial and final time moments, respectively; signs of $S_{\pm}$ correspond to the signs of noise $\eta(t)$ (\ref{cnoise}). Action $S_{\pm}$ corresponds to the potential $U(x)$ of the system if it is affected by Gaussian noise only, whereas action is different in the presence of non-Gaussian noise. The WKB approximation and action $S$  are being extensively applied for the analysis of fluctuations in non-equilibrium systems \cite{Luchinsky:98a,Khovanov:00,Luchinsky:02a,Luchinsky:02a}, whereas action $S$ specified a quasi-potential and has same meaning as potential in an equilibrium case. The latter means that the mean first-passage times $T_{\pm}$ can be presented in the following exponential form
\begin{equation}
\label{Kramers}
T_{\pm} \propto \exp \left( \frac{S_{\pm}}{\theta} \right)
\end{equation}
where $S_{\pm}$ is defined by (\ref{action}) with boundary conditions (\ref{bconds}). 

The theoretical approach presented above takes into account all cumulants of non-Gaussian noise. The third cumulant approximation can be obtained from equations (\ref{Ham_sys})  by expanding the exponential function into a series (\ref{expP}) and truncating all terms above $p_y^3$. The resulting Hamiltonian system is
\begin{eqnarray}
\label{Ham_sys3}
\begin{aligned}
\dot{x} &=y  \\
\dot{y} &= - \alpha y- \frac{d U}{d x} -\frac{\alpha D}{\theta}p_y +\frac{\lambda^2 \Lambda}{\theta} \left( -p_y \pm \frac{\lambda}{2\theta}p_y^2 \right)  \\
\dot{p}_x &= p_x \frac{d^2 U}{d x^2}  \\
\dot{p}_y &= -p_x+\alpha p_y  
\end{aligned} 
\end{eqnarray}
and corresponding action is defined as
\begin{eqnarray}
\label{action3}
S^{3}_{\pm}=&\int_{t_i}^{t_f} dt\left\{ -\frac{\alpha D}{2\theta}\tilde{p}_y^2-\frac{\lambda^2\Lambda}{2\theta}\tilde{p}_y^2\pm
\frac{\lambda^3\Lambda}{3\theta^2} \tilde{p}_y^3 
 \right\} \ 
\end{eqnarray}
where index ``3'' in (\ref{action3}) is used to indicate the third order cumulant approximation.
Thus, systems (\ref{Ham_sys}) and (\ref{Ham_sys3}) with boundary conditions (\ref{bconds}) and corresponding actions (\ref{action}) and (\ref{action3}) describe effects of the presence of non-Gaussian noise.
 We solve the boundary value problem using custom software~\cite{foot2}, following the approach described in paper~\cite{Beri:05}.

\section{Numerical simulations versus theory}
\label{sec:numerics}

Numerical simulations of the Langevin equation (\ref{sin_sys}) are extremely computationally demanding  because the escape time $T_{\pm}$ should be  lower than a characteristic relaxation time of the system (\ref{sin_sys}) by a factor $10^6$ for mimicking  the experiments \cite{Huard:07}. Therefore, for accelerating the simulations  we replace the periodic  potential $U(x)=\sin (x)-sx$ in (\ref{sin_sys}) by the third order polynomial $U(x)=ax-bx^3/3$ with parameters $a$ and $b$ to approximate  one well of the periodic potential (see Fig.~\ref{fig1}~(c)). Note that the relative difference between the theoretically calculated actions $S_{\pm}$ for these polynomial and periodic  potentials is less than 0.01\%. 

Numerical simulations were performed using a Heun difference scheme, details of which can be found in publication\cite{Khovanov:08}. The Poisson noise term was constant in each integration step and  was calculated \cite{Huard:07} as $\lambda \ {pn} (\Delta t \Lambda)$, where $\lambda$ and $\Lambda$ are the amplitude and frequency of the Poisson noise, respectively; $\Delta t$ is the integration step size and the value of $pn$ is produced by a pseudo-random numbers generator with a Poisson distribution (\ref{p_dist}). The applied scheme was verified against known theoretical results \cite{Malakhov_Book} for linear systems perturbed by non-Gaussian noise.  

First, we checked the scaling (\ref{Kramers1}) alongside the selection of the effective noise intensity as $\theta= \alpha D + \lambda^2 \Lambda$.
As it was mentioned above, our theoretical approach contains a self-contradiction: the asymptotic character of the WKB approximation aims to remove the explicit value of $\theta$, but  $\theta$ appears explicitly in the final expression. The same contradiction exists in previous theoretical developments \cite{Ankerhold:07,Huard:07,Sukhorukov:07,Grabert:08,Novotny:09} too, because $\theta$ includes  both Gaussian, $D$, and non-Gaussian,  $\lambda^2 \Lambda$, parts and this puts constraints on the way the parameter $\theta$ can be varied in numerical simulations to evaluate scaling (\ref{Kramers1}). As such the following factors have to be kept unchanged   
\begin{eqnarray}
\label{constrains}
\frac{\alpha D}{\theta}=C_1, \ \ \ \lambda \Lambda =C_2, \ \ \ \frac{\lambda}{\theta}=C_3, \ \ \ \theta\Lambda=C_4,
\end{eqnarray}
where $C_1$, $C_2$, $C_3$ and $C_4$ are constants, whereas the value of $\theta$ is varied. Importantly, these constraints keep the bias $I=\lambda \Lambda$ constant. The bias $I$ characterizes the asymmetry of the Poisson noise and it is the parameter of consideration \cite{Novotny:09}. The first condition in (\ref{constrains}) means that relative contributions of Gaussian and non-Gaussian noises are constant. Note that this is equivalent to keeping the ratio $\frac{\lambda^2 \Lambda}{\theta}$ constant; this ratio characterizes the relative contribution of Poisson noise. In order to calculate constants using  (\ref{constrains}), the values of bias $I=\lambda \Lambda$, ratio $\alpha D /\theta$ (or ratio $\frac{\lambda^2 \Lambda}{\theta}$) and $\lambda$ should be specified. Knowing the constants we can vary $\theta$ in order to change the values of $D$, $\lambda$ and $\Lambda$ as follows
\begin{eqnarray}
\label{way1}
\alpha D=C_1\theta, \ \ \ \lambda  =C_3 \theta, \ \ \ \Lambda=\frac{C_4}{\theta}.
\end{eqnarray}
It is evident that experimental implementation of the procedure described above is a non-trivial task; this is, however, the only means to  verify theoretical actions (\ref{action}) and (\ref{action3}). 
In the absence of Poisson noise these actions are equal to the potential barrier $\Delta U$
\begin{eqnarray}
\label{barier}
\Delta U&=& U(x_{max})-U(x_{min})=\frac{4}{3} \sqrt{\frac{a^3}{b}} 
\end{eqnarray}
with $x_{max}=\sqrt{a/b}$ and $x_{min}=-\sqrt{a/b}$ corresponding to the maximum and minimum of $U(x)=ax-bx^3/3$, respectively. For the selected values  $a=0.0093101$, $b=0.4965$ the potential barrier is $\Delta U=0.0017$.

\begin{figure}[t!]
\includegraphics[width=8cm]{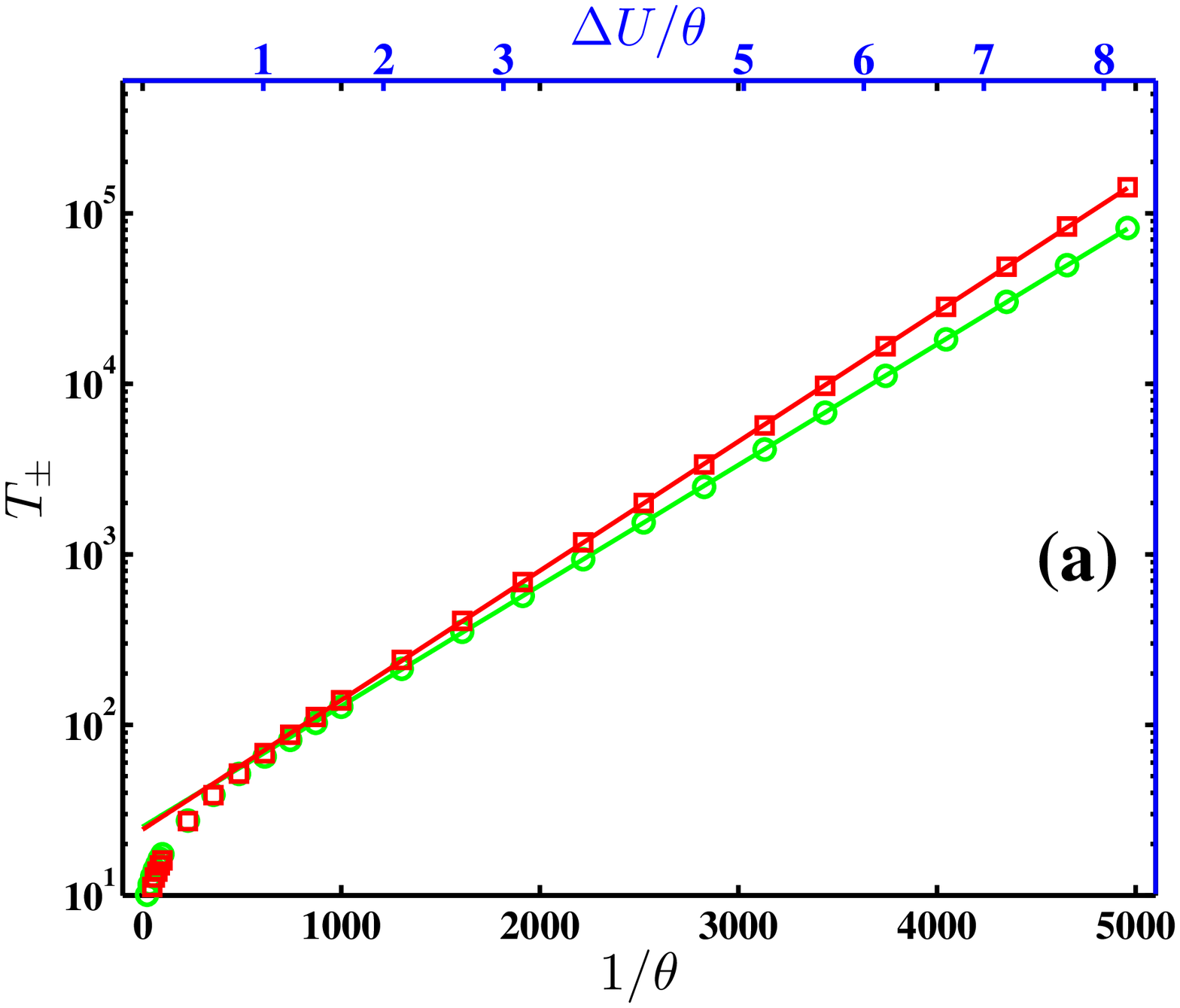}\\
\includegraphics[width=8cm]{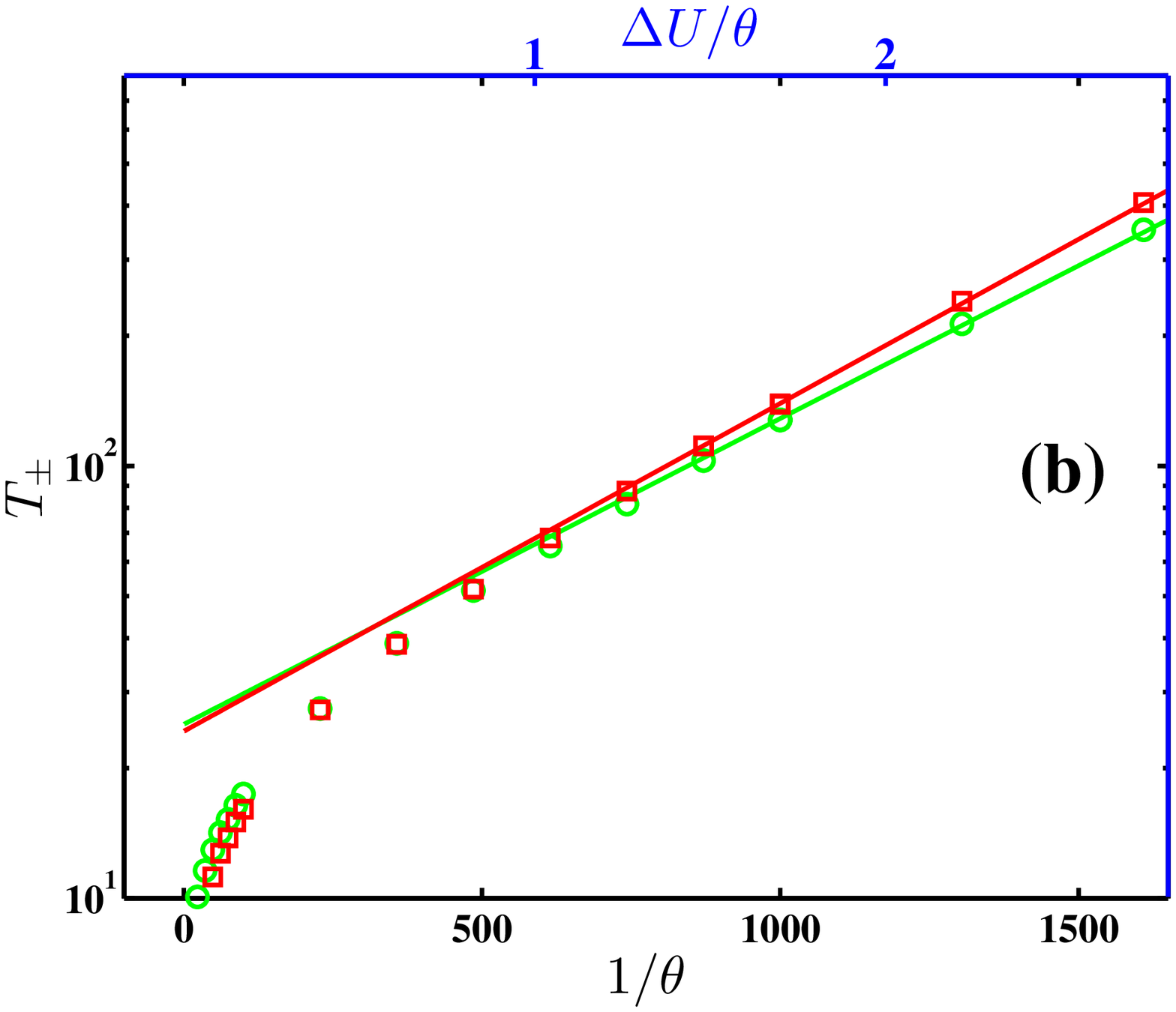}
\caption{\label{fig2} (color online) (a) Mean escape times $T_{\pm}$ as  functions of $1/ \theta$. Markers ``$\square$'' and ``$\circ$'' correspond to negative and positive signs of the term $\eta(t)$ in (\ref{sin_sys}) respectively. The calculations were performed for $\frac{\lambda^2\Lambda}{\theta}=0.3$ and $I=0.015$. Solid lines correspond to linear fitting using (\ref{lin_fit}). The scale of ordinate is logarithmic. (b) A zoomed part of figure (a). Upper abscissa in both figures shows values of the ratio $\Delta U / \theta $.
}
\end{figure}

Let us consider an under-damped regime (weak damping) of system (\ref{sin_sys}) by fixing the damping coefficient to $\alpha=0.5$; also, set ratio $\frac{\lambda^2 \Lambda}{\theta}=0.3$ and bias $I=0.015$. These values correspond to experimental conditions typically reported~\cite{Timofeev:07,Huard:07,Masne:09}. Following the approach described above we varied $\theta$ and calculated the mean escape times $T_{\pm}$ for the two different signs of the noise $\eta(t)$. More than $20\ 000$ escape events were used for the estimation of the values of $T_{\pm}$.
Escape times $T_{\pm}$, as functions of $1/\theta$, are shown in Fig.~\ref{fig2}. Exponential scaling (\ref{Kramers1}) is clearly observed and supports the validity of the selected form (\ref{expsc}) and the use of $\theta=\alpha D + \lambda^2 \Lambda$ as an effective noise intensity.  Significantly, the exponential scaling is observed in a wide range of $1/\theta$ up to the value of $\theta$  close to the magnitude of the potential barrier $\Delta U$, and therefore action $S$ (see upper axis in Fig.~\ref{fig2}~(b)). This result means that despite the asymptotic character $\theta \ll S$ of the  WKB approximation used, it is also applicable for $\theta \lesssim S$. This relaxes the condition constraining the use of a high barrier in experiments~\cite{Huard:07}. The range of experimental parameters for which the theoretical description is valid can therefore be significantly extended. 

The least-square fitting (solid lines in Fig.~\ref{fig2}) of the numerical results by a linear function 
\begin{eqnarray}
\label{lin_fit}
\log (T_{\pm})=S_{\pm} \frac{1}{\theta} + \log(Z_{\pm}),
\end{eqnarray}
allows us to extract both the values of actions $S_{\pm}$ and prefactors $Z_{\pm}$ in (\ref{Kramers1}) for conducting a comparison between the theory and numerical simulations.

\begin{figure}[t!]
\includegraphics[width=8cm]{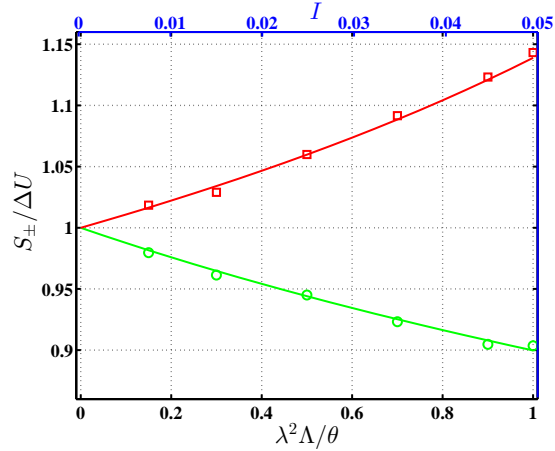}
\caption{\label{fig3} (color online) Normalized actions $S_{\pm}/ \Delta U$ are shown as  functions of the ratio $\frac{\lambda^2 \Lambda}{\theta}$.  Markers ``$\square$'' and ``$\circ$'' correspond to results of numerical simulations  for the  negative and positive signs of $\eta(t)$ respectively. Solid lines correspond to theoretical predictions. The value of $\theta=0.00042 $ was used in theoretical estimations of the actions $S_{\pm}$ by (\ref{action}). Other parameters are specified in the text. Upper abscissa shows value of the current $I$.
}
\end{figure}

Now let us vary $\alpha D$ from zero to $\theta$, that is  between the two extreme cases of pure Gaussian noise and pure Poisson noise. Mimicking experimental conditions, we fix the amplitude $\lambda=0.0084$ and vary frequency $\Lambda$. This results in varying the ratio $\frac{\lambda^2 \Lambda}{\theta}$ between 0 to 1. Note, that the bias $I$ also changes.
A comparison of numerical (markers) and theoretical (solid lines) normalized actions $S_{\pm}/\Delta U$ is presented in Fig.~\ref{fig3}. The results of the numerical simulations and  the theoretical predictions  are in close agreement thus proving the applicability of the theoretical approach presented here for the analysis of the non-Gaussian features of the noise. It is seen (Fig.~\ref{fig3}) that the difference between $S_{-}$ and $S_{+}$ increases with the increase of the relative contribution of Poisson noise which, in turn, corresponds to increased asymmetry  of noise distribution. The actual difference between the two values of $S_{-}$ and $S_{+}$ provides a qualitative description of the asymmetry.

\begin{figure}[t!]
\includegraphics[width=8cm]{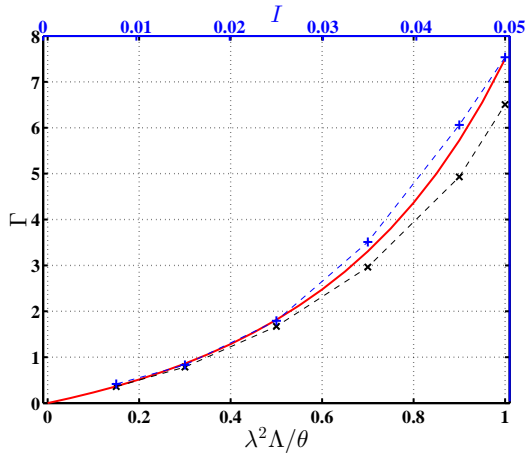}
\caption{\label{fig4} (color online) Asymmetry factor $\Gamma$ as  function of ratio $\frac{\lambda^2 \Lambda}{\theta}$. The solid curve corresponds to the theoretical factor $\Gamma_S$, markers ``$+$'' and ``$\times$'' correspond to $\Gamma_{S}^n$ and $\Gamma_T^n$ respectively. Parameters are selected as for Fig.~\ref{fig3}. Upper abscissa shows value of the current $I$.
}
\end{figure}

To reiterate, experimentally a non-Gaussian feature of noise is characterized by the asymmetry factor $\Gamma_T$ given by (\ref{Assymt}), whereas the theoretical approach estimates the factor $\Gamma_S$ via (\ref{Assym1}). Numerical simulations allow us to estimate both factors $\Gamma_T$ and $\Gamma_S$ by calculating the dependences $T_{\pm}(\theta)$ and using the fitting expression (\ref{lin_fit}) to extract the actions $S_{\pm}$. 
We denote numerically obtained factors (\ref{Assymt}) and (\ref{Assym1}) by a upper index $n$, that is  $\Gamma_T^n$ and $\Gamma_S^n$ respectively, and compare these factors with the theoretical $\Gamma_S$ for different values of the ratio $\lambda^2 \Lambda / \theta$.
As can be seen in Fig.~\ref{fig4}, the theoretical factor $\Gamma_S$  is close to $\Gamma_{S}^n$. This reflects once again the validity of the use of scaling (\ref{Kramers1}) as well as the applicability of the theory developed. The theoretical factor $\Gamma_S$ is also close to $\Gamma_T^n$ for small values of the ratio $\lambda^2 \Lambda / \theta$, when the asymmetry of the noise distribution is small. However, with this ratio  approaching 1, there is a growing difference between $\Gamma_T$ and $\Gamma_S$. The presence of such a difference  means that neglecting prefactor $Z$  in (\ref{Assym1}) can lead to an error when approximation (\ref{Assym1}) is used instead of factor (\ref{Assymt}) arising from experimental measurements. The maximum of the ratio $\lambda^2 \Lambda / \theta$ was reported to be around $0.6$ from experiments~\cite{Timofeev:07,Huard:07,Masne:09} and is within the range of negligible difference between $\Gamma_T$ and $\Gamma_S$. As a result, the use of $\Gamma_S$ in a theoretical consideration instead of the experimentally measured  $\Gamma_T$ cannot be a cause for a poor correspondence between theory and experiments~\cite{Huard:07,Novotny:09}. However, the ratio $\lambda^2 \Lambda / \theta$ can be larger in experiments~\cite{Pekola:05} and in this case the theory will produce an error in the estimation of the asymmetry factor $\Gamma_T$. 

\section{Comparison of the third order cumulant and all cumulants approaches}
\label{sec:3d}

\begin{figure}[t!]
\includegraphics[width=8cm]{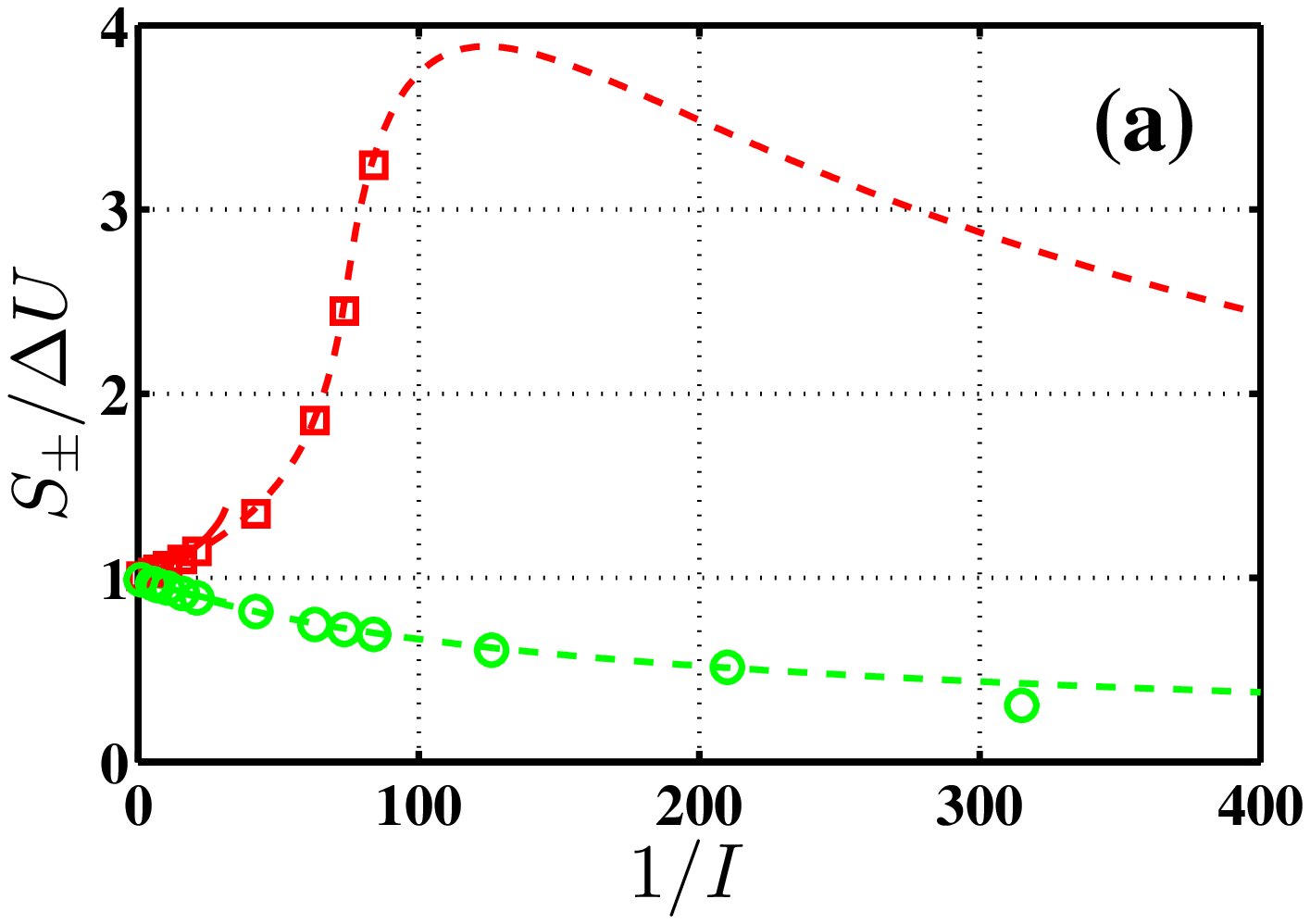} \\
\vspace{3mm}
\includegraphics[width=8cm]{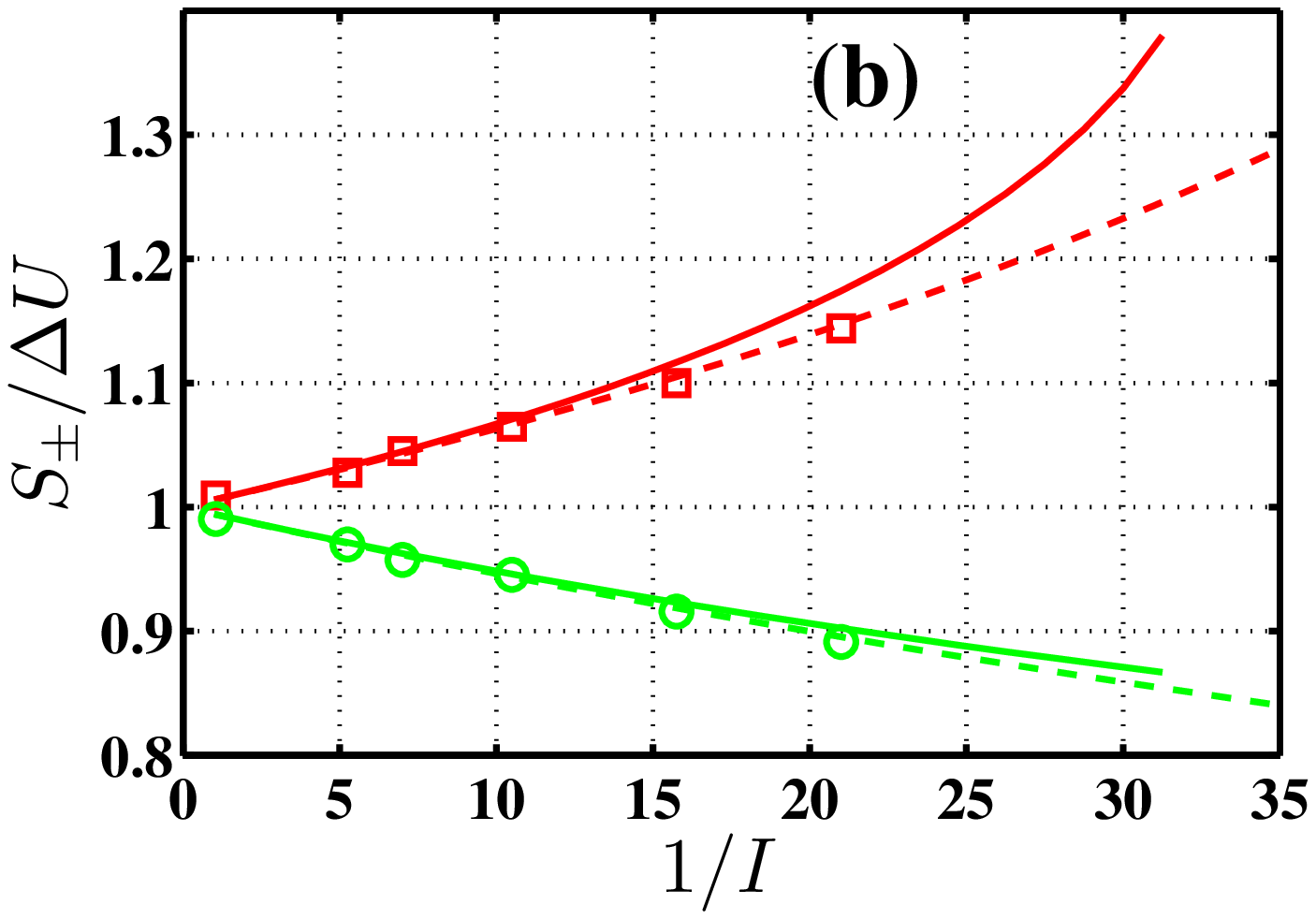} 
\caption{\label{fig5} (color online) Normalized actions $S_{\pm} / \Delta U$ as  functions of inverse current $1/I$.  Markers ``$\square$'' and ``$\circ$'' correspond to the results of numerical simulations  for negative and positive signs of $\eta(t)$ respectively. Dashed and solid curves correspond to actions $S_{\pm}$ (\ref{action}) and  $S_{\pm}^3$ (\ref{action3}). Parameters are specified in the text. Figures (a) and (b) show dependences $S_{\pm}(1/I)$ for different ranges of $1/I$. 
}
\end{figure}

Previous theoretical developments \cite{Ankerhold:07,Huard:07,Sukhorukov:07,Grabert:08,Novotny:09} aimed to derive an analytical expression with the third order cumulant approximation.  In this section, the importance of keeping all cumulants is considered  via a comparison of two actions $S_{\pm}$  and $S_{\pm}^3$ calculated according to  (\ref{action}) and (\ref{action3}), respectively, with the numerically obtained action via scaling (\ref{Kramers1}). For maximizing  the effects of non-Gaussianity of noise, the pure Poisson noise has been investigated, i.e. when $D=0$.  The bias $I=\lambda \Lambda$ is selected as a varying parameter characterizing the asymmetry of the noise distribution. Non-Gaussian effects are maximized in the $I\rightarrow 0$ limit; another limit as $I\rightarrow \infty$ corresponds to the Gaussian case. 
The dependences of the theoretical and numerical actions as  functions of the inverse current $1/I$ are shown in Fig.~\ref{fig5}. Several remarkable features can be seen.

Firstly, the third cumulant approach provides the solution for the limited range of the inverse current $1/I<32$ only. Outside this range, the solution of the boundary value problem (\ref{bconds}) does not exist for negative $\eta(t)$, whereas this is not the case for the positive sign.
For large values of bias $I$ ($I>0.2$ or for the inverse $1/I<5$),  the difference between the all cumulants and third cumulant approaches (Fig.~\ref{fig5}, {\it b}) is small, however  the difference increases with the decrease of the bias $I$  (greater $1/I$). Note that in the experiments \cite{Huard:07} the current is relatively large and consequently, the third cumulant approach provides high accuracy predictions.

Secondly, for the all cumulants approach, there is a very good correspondence between theoretical and numerical results for a wide range of $1/I$. For $1/I >90$ (Fig.~\ref{fig5}, {\it a}) the exponential scaling (\ref{Kramers}) is not observed in numerical dependencies $T(\theta)$ for the negative sign of $\eta(t)$. 

Thirdly, the theory predicts a bell-shape of the dependence $S_{-}(1/I)$  (red dashed line in Fig.~\ref{fig5}(a)) with a clear maximum. This feature was not confirmed by numerical simulations for the selected parameters, but it was observed for a different set of parameters. Further discussion of this feature is out of the scope of the current manuscript. 

\begin{figure}[t!]
\includegraphics[width=8cm]{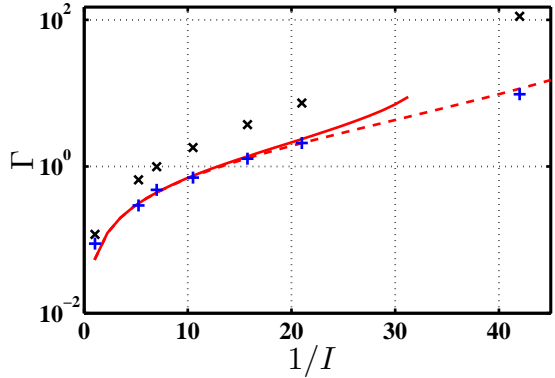}
\caption{\label{fig6} (color online) Asymmetry factors $\Gamma$ as  functions of inverse current $1/I$. Solid and dashed curves correspond to theoretical factors $\Gamma_{S}^3$ and $\Gamma_S$ respectively. Markers ``$+$'' and ``$\times$'' correspond to $\Gamma_{S}^n$ and $\Gamma_T^n$ respectively.
}
\end{figure}

Finally, numerical and theoretical asymmetrical factors $\Gamma$ calculated as functions of the inverse bias $1/I$ (Fig.~\ref{fig6}) have been compared.  
There is an excellent correspondence between the numerical factor $\Gamma_S^n$ and the theoretical factor $\Gamma_S$ which was obtained by the all cumulants approach. The third order cumulant approximation $\Gamma_S^3$ has a limited range of $1/I$ where the theoretical prediction is close to numerical results. The difference between the factors $\Gamma_S$ and $\Gamma_T$ is observed in a wide range of the inverse bias $1/I$ and moreover, the difference reaches a value of one order of magnitude. The latter demonstrates a significant contribution of the prefactors $Z_{\pm}$ in the estimation of the asymmetry factor $\Gamma_T$ for the case of strong asymmetry of noise distribution. 

\section{Conclusions and discussion}
\label{sec:concs}

We presented a theoretical background for calculating the action (an exponential factor of the mean escape time) for an under-damped oscillator driven by a mixture of white Gaussian and Poisson noises. Note that this approach can be extended to systems of any dimension and to any non-Gaussian noise with finite cumulants. The validity of the theoretical approach suggested here has been confirmed by numerical simulations. We showed that this approach is able to provide a qualitative prediction for actions $S_{\pm}$ for a wide range of parameters.

Theoretical considerations as presented here and published elsewhere~\cite{Pekola:04,Ankerhold:07,Grabert:08,Novotny:09,Huard:07} include the asymptotic parameter $\theta$ explicitly in the final expressions. This places constraints  (\ref{constrains}) on the range of the parameters in experiments or numerical simulations within which they can be varied in order to be consistent with the theoretical approach. This implies that the dependence of the mean first passage time $T$  on noise intensity is not exponential  (\ref{Kramers}) for the case where intensity of only one noisy component in a mixture of two components (Gaussian and non-Gaussian) is varied. Note that the exponential scaling of escape rate as a function of intensity of the Poissonian component (Gaussian component (temperature) was constant) was used for a comparison between the theory and experiments~\cite{Huard:07,Masne:09}. According to our results (Fig.~\ref{fig3}) varying the intensity of just one component changes the ratio between the components and  it changes action $S$, which is represented by an exponential factor in $T$. Since this change of $S$ is relatively small, the deviation of $T$ from exponential scaling is weak but still present in experiments~\cite{Huard:07,Masne:09}. It is noticeable that there is a difference  between experimental and theoretical results (for example, compare dashed and solid lines in Fig.~2~(b) in Le Masne et al~\cite{Masne:09}); this difference can be explained by the action being dependent on the intensity ratios between Gaussian and Poissonian components of noise. 

It was stated in the introduction that there are discrepancies between the published theoretical predictions and experiments. As a possible explanation~\cite{Novotny:09} of the discrepancies the use of a third order cumulant approximation and omission of a prefactor in the escape rate have been mentioned. Our comparative study shows that these are not relevant for the range of parameters in experiments. The discrepancy can be explained by the additional approximations made in equation (\ref{action3}) necessary for deriving an analytical expression for the action $S$. In contrast, we solved the equation {\it numerically}, and showed excellent correspondence  between the theoretical approach and numerical simulations thus confirming the validity of the general theoretical framework.   
 
We showed that the third order cumulant approach is applicable for a limited region of parameters, within which the non-Gaussian effects are relatively weak. The all cumulants approach does not have such limitations and demonstrates an excellent correspondence with the results of numerical simulations. We further demonstrated that both the current and all previous theoretical approaches \cite{Ankerhold:07,Huard:07,Sukhorukov:07,Grabert:08,Novotny:09} are not capable of providing a {\it quantitative} description of noise with a {\it strong} asymmetry since these do not take the prefactor $Z_{\pm}$ into account in the expression for the mean escape rate (\ref{Kramers1}) and subsequently in (\ref{Assymt}) for the asymmetry factor $\Gamma_T$. These approaches can only provide a {\it qualitative} prediction. Furthermore, since the theory is able to accurately predict the actions $S_{\pm}$, experiments need to be designed so as to extract the actions rather than the mean escape time. Such experiments would then lead to new challenges as these will require simultaneous tuning of several parameters in order to satisfy four conditions (\ref{constrains}).

\section{Acknowledgments}

The authors gratefully acknowledge T. Novotny for valuable discussions. The authors would like to thank N. Evans and the anonymous reviewers for their constructive comments and suggestions that led to the improvement of the original version of this article. 
The work has been supported by the EPSRC (EP/C53932X/2,  EP/G070660/1 and EP/K02504X/1).


\begin{thebibliography}{45}
\bibitem[{foo()}]{foot2}
\bibinfo{note}{Although we have tried to write code usable by a third party,
  it still requires improvement. Nevertheless, we are happy to share the code
  upon request.}
\end{thebibliography}

\end{document}